\documentstyle[buckow1,12pt,epsf,fps]{article}
\newcommand{\Z}{Z \!\!\! Z}
\begin {document}
\def\titleline{
Perfect and Quasi-Perfect Lattice Actions
}
\def\authors{
W. Bietenholz \1ad
}
\def\addresses{
\1ad
HLRZ c/o Forschungszentrum J\"{u}lich, D-52425 J\"{u}lich, Germany\\
} \vspace*{-2mm}
\def\abstracttext{
Perfect lattice actions are exiting with several respects:
they provide new insight into conceptual questions of the
lattice regularization, and quasi-perfect actions could enable
a great leap forward in the non-perturbative solution of QCD.
We try to transmit a flavor of them, also beyond the lattice community.
}
\makefront

\vspace*{-10mm}
\section{Introduction}

For many field theoretic models, in particular in $d=4$,
the lattice is the only regularization so far, which provides
non-perturbative results.
One discretizes the Euclidean space and 
introduces matter fields only on the lattice sites, and
gauge fields on the links connecting them.
By means of some Monte Carlo procedure one generates a number
of configurations, and for large statistics
we can numerically perform a functional integral (to some accuracy).
However, even if the statistical error is under control,
we still have to worry about systematic errors.
Simulations take place at a finite lattice spacing
$a$ and in a finite size $L$, so the final limits
$\bar \xi /a , \ L/\bar \xi \to \infty$ require extrapolations of
the simulation results ($\bar \xi$ is the correlation length).
The finiteness of these ratios causes artifacts, and in 
practice those due to $a>0$ are most troublesome. 

It is very expensive in computer time to approximate
the continuum limit by ``brute force'', i.e. by using finer and finer
lattices. The computational effort grows at least like $a^{-6}$,
in some cases this factor can rise up to $a^{-10}$. 
When he was about to quit lattice physics, K. Wilson
estimated that a convincing solution of QCD requires a lattice
of $256^{3}\times 512$ sites. This appears hopeless for generations
of supercomputers as well as human beings; what the up-coming
supercomputers (``teraflops'')
can process is around $24^{4} \dots 32^{4}$ for full QCD.
(Larger lattices can be used in the so-called quenched approximation:
there one sets the fermion determinant equal to a constant,
which means physically that sea quark effect are 
neglected. Hence one brings in another systematic error, which
is often difficult to control.)

However, Wilson referred to his standard lattice action,
which  handles derivatives by nearest lattice site differences
(including a possible gauge variable living on the connecting link).
The pure gauge part is described by a ``plaquette action'':
from the paths around single plaquettes one constructs
a (gauge invariant) lattice version of $F_{\mu \nu}$ \cite{books}.
Nowadays, there is a consensus that
{\em improved lattice actions} are a ray of hope
for a much faster progress in the near future, than 
the gradual increase in computer power provides.
Of course, many lattice actions have the correct continuum
limit, and a better choice could suppress the artifacts
significantly. Non-standard lattice actions involve additional terms,
which make the simulations more complicated and slower. Still
the ratio gain/cost can be huge, if they allow us to, say,
double or triple the lattice spacing, which
has to be chosen typically $\leq 0.1 fm$ for Wilson's QCD action.

At present, there are essentially two improvement programs
under investigation. One of them has been formulated by K. Symanzik
\cite{Sym}, and tries to cancel the cutoff artifacts 
order by order in $a$,
by adding irrelevant operators to Wilson's action. This is
very similar to the Runge and Kutta procedure for solving
differential equations. In QCD 
(with Wilson fermions) it has been realized to
$O(a)$, first on the classical level by adding
a so-called ``clover term'' (a plaquette version of 
$\sigma_{\mu \nu} F_{\mu \nu}$) \cite{SW}.
More recently, the renormalization of the clover coefficient has 
been estimated by a mean-field approach \cite{LepMac}, and finally the 
non-perturbative $O(a)$ improvement has been completed by extensive 
simulations, taking the PCAC relation as a guide-line \cite{ALPHA}.
At present, many tests are being performed with the resulting
action, but it is not clear yet to which extent it enables the
use of coarse lattices \cite{Hors}. 
As a preliminary impression, there is progress in the scaling
on fine lattices, but it does not enable the use of really coarse
lattices (it seems that this $O(a)$ improvement accidently amplifies
the $O(a^{2})$ artifacts).

These notes are devoted to the alternative improvement program,
which goes under the name ``perfect actions''.
It is based on renormalization group concepts, and it is
non-perturbative with respect to $a$.
It works beautifully in principle,
as we know from a sequence of 2d toy models:
the $O(3)$ model, the Gross-Neveu model, the $CP(3)$ model,
and the Schwinger model.
Moreover, it has theoretically fascinating properties.
In particular, it allows us to reproduce
symmetries of the continuum theory exactly on the lattice,
even in cases where this is apparently impossible.
Examples are the continuous Poincar\'{e} invariance, as well
as chiral symmetry. It is even possible to introduce a 
perfect lattice topology.
Hence the program is very attractive from the conceptual point
of view, also apart from the practical aspect of reducing
the computational effort in simulations.

However, the implementation in full QCD is very tedious.
In practice, a crucial aspect is the struggle for an excellent 
locality of the action, so that the truncation of the couplings
-- which is required at some point -- does not do too much harm to it. 
In this context, a good parameterization
and truncation of the action are now major issues in this
program, which have not yet been solved in a really satisfactory way.
Still, the potential of this method is great; if it can really
be applied, then it can do by far better than an $O(a)$
improvement, although it requires more non-standard terms than
just a clover term. In this program, the improvement
can be extended to better and better approximations
to perfection. In contrast, it is hardly feasible
to carry on Symanzik's program to $O(a^{2})$, so if
$O(a)$ should be insufficient with some respect, 
then that program is at a dead-lock.

\section{Perfect and classically perfect actions}

It has been known for a long time that there exist so-called
{\em renormalized trajectories} in parameter space \cite{WilKog},
\footnote{One may think of the parameter space as being spanned
by all possible couplings between the lattice variables.}
which represent
{\em perfect lattice actions}. These are actions without any 
lattice artifacts; they display the continuum values of 
scaling quantities at any lattice spacing.

This can be understood from the consideration of block variable
renormalization group transformations (RGTs).
As a simple example, we start from a hypercubic lattice with 
spacing $1/n$, and divide it into
disjoint blocks of $n^{d}$ sites, where $d$ is the Euclidean
space-time dimension, and $n$ is going to be the blocking factor.
The block centers form a coarse lattice of unit spacing. There
we define new variables, collectively denoted by $\phi '$,
which we relate to the corresponding block averages of the
fine lattice variables $\phi$. The action on the coarse lattice
is given by
\begin{equation}
e^{-S'[\phi ']} = \int D\phi \ K[\phi ',\phi ] e^{-S[\phi ]} .
\end{equation}
The kernel $K$ must be chosen such that the partition
function, and all expectation values -- hence the physical
contents of the theory -- remain invariant.
This requires
\begin{equation}
\int D\phi ' \ K[\phi ' , \phi ] = 1,
\end{equation}
which still leaves quite some freedom.
The simplest choice,
\begin{equation} \label{deltaRGT}
K [ \phi ' , \phi ] = \prod_{x'} \delta (\phi_{x'}'-\frac{1}{n}
\sum_{x\in x'} \phi_{x}) ,
\end{equation}
determines the $\delta$ function RGT (the sum $x\in x'$ runs over
the fine lattice sites $x$ in the block with center $x'\in \Z^{d}$).
An obvious generalization of the kernel (\ref{deltaRGT})
``smears'' the $\delta$ function to a Gaussian, so that
the transformation term appears in the exponent.
The RGT reduces the correlation length {\em in lattice units},
$\xi $, by a factor $n$, $\xi ' = \xi /n$.

Assume that we are on a ``critical surface'' in parameter space,
where $\xi = \infty$. For suitable parameters
we arrive -- after an infinite number of RGT iterations 
(which include permanent re-scaling to the newest lattice units) --
at a finite fixed point action
(FPA) $S^{*}$, which is invariant under the considered 
RGT: $S^{*}{'}=S^{*}$.

Let us now perform a tiny step away from the FPA 
-- and from the critical surface -- in a relevant
direction, and then keep on iterating RGTs. Thus we follow a trajectory
in parameter space, which takes us to shorter and shorter correlation 
length. This is a renormalized trajectory, each point of which is 
related to the vicinity of the FPA solely by the renormalization group. 
There is no way irrelevant operators can contaminate the actions
represented by the points on such a trajectory, hence these actions
are perfect. Scaling quantities extracted from such actions are 
therefore completely free of lattice spacing artifacts.
Needless to say that the identification of (quasi-)perfect
actions at moderate or even short $\xi$ -- where the simulations
take place -- is a dream of humanity (or at least of the lattice
community). However, it is very difficult to find good
approximations to perfect actions, which are tractable in simulations.

As an alternative description, one may start on a suitable 
point close to the critical surface, and after many RGTs
a renormalized trajectory is approximated asymptotically.
This property suggests an explicit construction by Monte Carlo
RGTs, which has been tried for some time, but which did not
prove very useful: such RGT steps are tedious to perform,
and the iteration is restricted to very few (often hardly one) 
reliable steps, which do not suppress the artifacts 
dramatically.

\begin{figure}[hbt]
\vspace{-0.2cm}
\hspace*{0.5cm}
\def\fpsangle{270}
\epsfxsize=70mm
\fpsbox{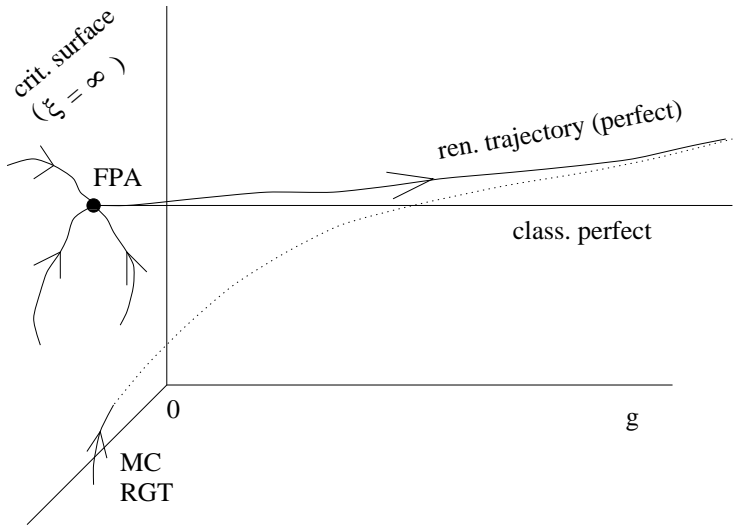}
\vspace{-0.5cm}
  \caption{\footnotesize
            {A caricature of the picture in parameter space.}}
   \label{rgtfig}
\end{figure}

\begin{figure}[hbt]
\vspace{-0.5cm}
\hspace{3.5cm}
      \epsfxsize=8.00cm \epsfbox{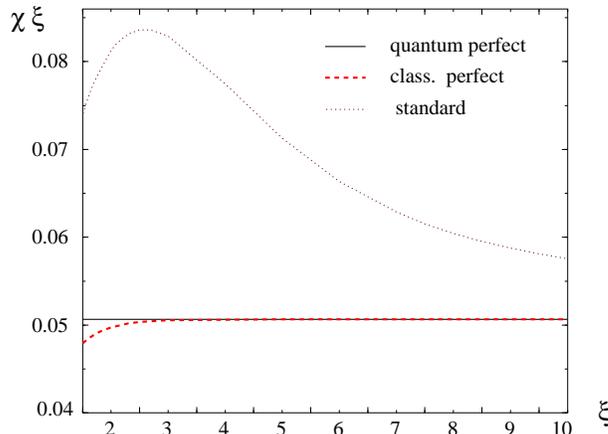}
\vspace{-0.5cm}
   \caption{\footnotesize
            {The scaling of the topological susceptibility $\chi$
(multiplied by the correlation length in lattice units, $\xi$) 
in the 1d XY model for the (quantum) perfect action, the classically 
perfect action and the standard lattice action.}}
   \label{topscal}
\end{figure}

A few years ago, P. Hasenfratz and F. Niedermayer suggested a new
trick to construct approximately perfect actions, which is
particularly designed for asymptotically free theories \cite{HN}.
We write the RGT (in a slightly modified notation) as
\begin{equation}
e^{-\frac{1}{g{'}^{2}} S'[\phi ']} = \int D \phi \
e^{-\frac{1}{g^{2}} \{ S[\phi ] + T [\phi ' ,\phi ] \} } ,
\end{equation}
where $T$ is now the transformation term, which specifies the RGT. 
Here the critical surface -- and hence the FPA --
is situated at $g=g'=0$, where the RGT is determined simply
by minimization,
\begin{equation} \label{mini}
S'[\phi ' ] = \ ^{min}_{~\phi} \{ S[\phi ] + 
T[\phi ' , \phi ]\} , \quad
S^{*}[\phi '] = \ ^{min}_{~\phi} \{ S^{*}[\phi ] + 
T[\phi ' , \phi ]\} .
\end{equation}
Thus the search for a FPA simplifies enormously to a 
classical field theory problem; 
no (numeric) functional integral is needed.
\footnote{In the literature, this is often called a
``saddle point problem'', although one just deals with minima.}
\footnote{In practice one may proceed as follows:
make a parameterization ansatz for the action $S'$;
choose some configurations $\phi '$; (numeric) minimization
yields $S'[\phi ']$. For a sufficient number of configurations,
the parameters in the ansatz for $S'$ are determined.}
But we still need to proceed to moderate $\xi$ in order
to control the finite size effects. 
And here the authors of Ref. \cite{HN} suggest to just switch on
$g$ and use the ``{\em classically perfect action}'' 
$(1/g^{2}) S^{*}[\phi ]$. Thus we follow the weakly relevant
(in leading order marginal) direction away from the FPA.
If we multiply $g^{2}$ by $\hbar$, we understand the notion
``classically perfect''; in the limit $\hbar \to 0$
the minimization trick persists at finite $g$.
In the full quantum theory, however, this is not the case;
the renormalized trajectory deviates from its classically
perfect approximation. The hope -- and the one uncontrolled assumption 
in this program -- is that the latter is still an approximately
perfect action at moderate $\xi$. (Intuitively we may 
support this hope by associating the quantum
deviations of renormalized trajectories with the $\beta$ functions,
which tend to be smooth for asymptotically free theories.)
In fact, toy model studies confirm the excellent
quality of this approximation
in a striking way: in a scaling test for the 2d $O(3)$ model
(in a small volume $L\simeq \xi$)
no lattice artifacts at all could be seen down to $\xi \simeq 5$.
A study of the Gross-Neveu model revealed that in the
large $N$ limit the classically perfect action is also
quantum perfect ($N$ suppresses the quantum fluctuations)
so that, for instance, the dynamically generated fermion mass
divided by the chiral condensate is a constant, independent
of $\xi$ \cite{GN}.
The classically perfect 2d $O(3)$ action, together with a classically 
perfect topological charge, confirmed accurately
the absence of scaling of the topological susceptibility \cite{O(3)topo}.
Decent topological scaling was found, however, using a classically 
perfect action for the 2d $CP(3)$ model \cite{CP(3)}.
In the 1d XY model even a quantum perfect topology was worked out:
for a given lattice configuration, we integrate over all possible
continuum interpolations, each with a well defined topological
charge. Thus we attach to a lattice configuration an ensemble 
of charges with appropriate Boltzmann weights \cite{XY}. Furthermore,
the scaling artifacts of the classically perfect approximation
could be studied: they become negligible around
$\xi \simeq 3$, whereas the standard lattice formulation still suffers
from severe artifacts at large $\xi$, see Fig. \ref{topscal}.

In the Schwinger model with Wilson-type fermions,
the classically perfect action was again very successful,
in particular in view of the scaling of ``meson'' masses;
also the dispersion relations and the rotational invariance
of the correlation functions look good \cite{Lang}.
Approximations to FPAs were also constructed for non-Abelian 
gauge fields in $d=4$. 
They have been applied to studies of topology of $SU(2)$ 
\cite{SU(2)}, and various types of strongly truncated FPAs
have been suggested for $SU(3)$ \cite{SU(3)} (but recently 
those authors reported certain problems with their actions).

Let us  emphasize again that the improvement is 
designed to improve the {\em scaling}, i.e. scaling quantities
(dimensionless ratios of observables) should converge
to their continuum limit at smaller $\xi$ than it is the
case for the standard action. The impact on {\em asymptotic
scaling} can not be predicted.
However, it has been observed that also asymptotic scaling tends 
to set in much earlier for quasi-perfect actions 
\cite{GN,SU(3),anos}. In particular,
$\Lambda_{QCD}$ is much larger -- hence much closer to its
continuum value -- for the classically perfect action
(compared to Wilson's action) \cite{Bielef}.

\subsection{Classically perfect operators}

For a given configuration $\Phi $ on the coarse lattice,
the first eq. in (\ref{mini}) singles out one configuration
$\phi_{c}$ on the fine lattice, which minimizes the
right-hand side. This can be iterated to finer and finer 
interpolating lattices (in units of the original coarse lattice), 
until we arrive at a minimizing continuum field $\varphi_{c}[\Phi ]$,
which we call the ``classically perfect field'' \cite{QuaGlu}.
It can be viewed as a particularly
sophisticated interpolation of the initial lattice field.
In contrast to ordinary interpolations, this inverse blocking
process is based on the renormalization group, at least in
the classical limit, and its artifacts tend to be suppressed
exponentially.
It can be used to define classically perfect
topological objects on the lattice, by requiring their stability 
under inverse blocking. This implies a better foundation
for long range stability than
just smoothing the lattice field locally by hand.

Now consider an operator ${\cal O} [\varphi ]$ given as a functional
of the continuum fields $\varphi$. We build a classically perfect
version of such operators simply by the substitution
\begin{equation}
{\cal O}[\varphi ] \to {\cal O}[\varphi_{c}[\Phi ] ] .
\end{equation}
Thus the operator is given in terms of lattice fields,
but still defined in the continuum, again as a sophisticated 
interpolation.

\begin{figure}[hbt]
\hspace*{3.5cm}
\def\fpsangle{270}
\epsfxsize=60mm
\fpsbox{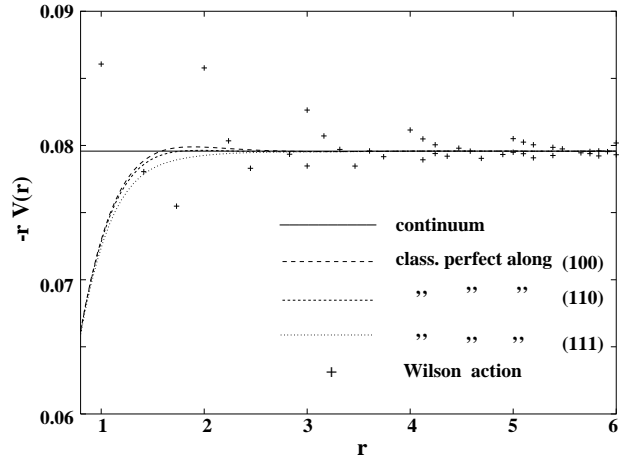}
  \caption{\footnotesize
            {The static quark-antiquark potential
(multiplied by the negative distance $-r$) obtained from 
the classically perfect action and from the plaquette action. }}
   \label{pot}
\end{figure}

As an application, this procedure has been applied to the
free gauge field, referring to a lattice with unit 
spacing \cite{QuaGlu}. From the classically perfect field,
we constructed Polyakov loops, the correlation function
of which yields a static quark-antiquark potential $V(\vec r )$.
Indeed, the classically perfect potential converges to the 
continuum value with increasing $r=\vert \vec r \vert$ 
much faster than the potential arising from Wilson's
plaquette action, it is continuously defined and 
-- most importantly -- it has an amazing
degree of rotational invariance, even at very short distances,
see Fig. \ref{pot}.

\section{Perturbatively perfect actions}

As another approximation to perfection, perfect actions can be
calculated analytically for free and perturbatively
interacting fields.
As potential applications, one
can hope that such actions are immediately useful in
simulations, or -- more modestly -- that they single out the
most important non-standard lattice terms, which gives a
handle for a good parameterization.
Moreover, they provide a promising starting
point for the minimization program, leading to a classically
perfect action. This is important, because even the
minimization can not easily be iterated.

\subsection{Free fermion and gauge fields}

In cases where the blocking step can be performed analytically,
it is more efficient to send the blocking factor $n\to \infty$
and to perform only one step, which yields the perfect
action directly, instead of tedious iterations.
Hence (in coarse lattice units) the
initial action lives in the continuum, so we don't need
to worry either which lattice action to use on the fine lattice.

Consider the free fermion. As described in Sec. 2, one usually
relates $\psi_{x}' \sim (1/n^{d}) \sum_{x'\in x} \psi_{x}$,
where $\psi ', \psi $ are fermionic fields 
on the coarse resp. fine lattice.
In the limit $n \to \infty$ this relation turns into
$\Psi_{x} \sim \int_{C_{x}} dy \ \psi (y)$, where 
$\Psi_{x}$ lives on a unit lattice, $C_{x}$
is the unit hypercube with center $x$, and $\psi$ is a continuum
field. In momentum space, this relation reads
\begin{equation} \label{blocking}
\Psi (p) \sim \sum_{l\in \Z^{d}} \psi (p+2\pi l) \Pi (p+2\pi l),
\quad \Pi (p) = \prod_{\mu} \frac{\hat p_{\mu}}{p_{\mu}}, \quad
\hat p_{\mu} = 2 \sin \frac{p_{\mu}}{2} ,
\end{equation}
where $p \in B = ]-\pi ,\pi ]^{d}$. If we compute the Gaussian type
RGT 
\footnote{We ignore constant factors in the partition function.
The RGT parameter $\alpha$ is arbitrary. The mass $m$ is given
in lattice units, even in the continuum action.}
\begin{eqnarray}
e^{-S[\bar \Psi , \Psi ]} &=& \int D\bar \psi D \psi
\exp \Big\{ - \int \frac{dp}{(2\pi )^{d}} \ \bar \psi (-p)
[i p \!\!\! \slash  + m] \psi (p) \nonumber \\
&& - \frac{1}{\alpha} \int_{B} \frac{dp}{(2\pi )^{d}} \
[\bar \Psi (-p) - \sum_{l\in \Z^{d}} 
\bar \psi (-p-2\pi l) \Pi (p+2\pi l)]
\times \nonumber \\ \label{fermiRGT}
&& \qquad \qquad \qquad
[\Psi (p) - \sum_{l\in \Z^{d}} 
\psi (p+2\pi l) \Pi (p+2\pi l)] \ \Big\} \ ,
\end{eqnarray}
we obtain the perfect lattice action
\begin{equation} \label{perfact}
S[\bar \Psi , \Psi ] = \int_{B} \frac{dp}{(2\pi )^{d}}
\bar \Psi (-p) G(p)^{-1} \Psi (p) , \quad
G(p) = \sum_{l\in \Z^{d}} 
\frac{\Pi^{2}(p+2\pi l)}{i(p_{\mu}+2\pi l_{\mu})
\gamma_{\mu} + m} + \alpha ,
\end{equation}
which we write in coordinate space as
\begin{equation}
S[\bar \Psi , \Psi ] = \sum_{x,r \in \Z^{d}} \bar \Psi_{x}
[\rho_{\mu}(r) \gamma_{\mu} + \lambda (r)] \Psi_{x+r}.
\end{equation}
The couplings $\rho_{\mu}(r),\ \lambda (r)$ have been evaluated 
numerically \cite{QuaGlu}.

\begin{figure}[hbt]
\def\fpsangle{270}
\epsfxsize=42mm
\fpsbox{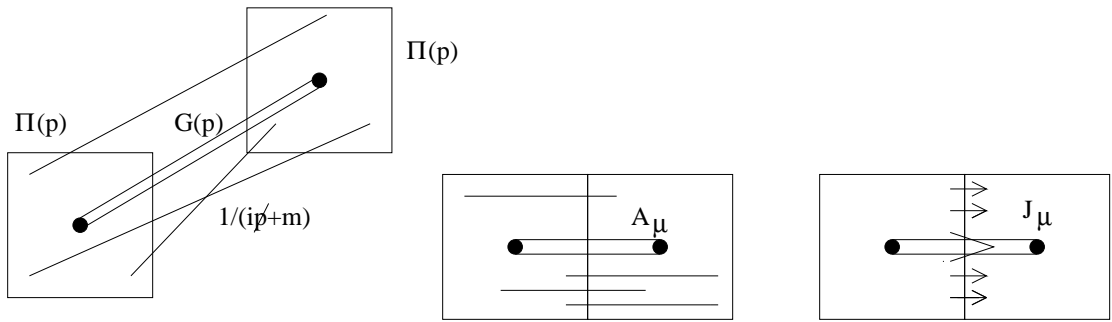}
  \caption{\footnotesize
            {The blocking scheme for the free fermion
propagator, for the  Abelian gauge field, and for any sort of currents.}}
   \label{block}
\end{figure}

The spectrum of this perfect lattice fermion can be read off
from the pole structure of the propagator $G(p)$,
\begin{equation}
E(\vec p \, )^{2} = (\vec p + 2\pi \vec l \, \, )^{2}+m^{2},
\end{equation}
hence it is the {\em exact} continuum spectrum (plus $2\pi$
periodic copies, which are omnipresent on the lattice).
This reveals the perfect character of this action:
the continuous rotation and translation invariance
is exactly present {\em in the observables}, though not in
the form of the action itself. In the latter, the
hypercubic structure of the lattice is visible -- a perfect action on a
triangular lattice, for instance, looks different \cite{WB}
-- but what matters are the observables.\\

The procedure described above, which we call ``blocking from the 
continuum'', is also applicable to the Abelian gauge field.
Here we integrate over all straight connections between
corresponding continuum points in two adjacent lattice cells, and
relate this integral to the non-compact lattice link
variable $A_{\mu}$. (A RGT in terms of ``compact'' link 
variables $U_{\mu} \in U(1)$ in $d=2$ has been carried out in
Ref. \cite{BenGro}.)

\subsection{Chiral symmetry}

A major issue in fermionic actions is the doubling problem:
according to the Nielsen Ninomiya ``No Go theorem'', species doubling
(unphysical extra fermions) always occurs under some mild assumptions
about a lattice action,
like locality, hermiticity and chiral invariance \cite{NN}.
For instance, in the naive fermion formulation with the inverse
lattice propagator $G_{naive}^{-1}(p)= i \sin p_{\mu}\gamma_{\mu}$
(at $m=0$), this is obvious from the occurrence of $2^{d}$ zeros in
the first Brillouin zone $B$. It is a notorious problem to
put a single chiral fermion on the lattice.

It turns out that the perfect action (\ref{perfact}) is not
plagued by doubling. Moreover, in the case $m=\alpha=0$, 
\footnote{The case $\alpha=0$ corresponds to the $\delta$ 
function RGT.}
the action is chirally symmetric, but in this case
it is non-local ($\vert \rho_{\mu}(r)\vert $ only decays like
$\vert r \vert ^{1-d} $ at large distances)
so there is no contradiction with the No Go theorem \cite{nonloc}. 
As soon as $m> 0$ or $\alpha > 0$ (or both), the action becomes
{\em local}, i.e. the couplings in $\rho_{\mu}$ and $\lambda$
decay exponentially in $\vert r \vert $, and at the same time the
chiral symmetry is explicitly broken in the action.

But this is {\em not} the end of the story. If we start from a chiral
fermion in the continuum, then the chiral symmetry is supposed to
be preserved under the RGT, due to its very nature, also for finite
$\alpha$. And this is in fact the case, as we see if we focus on the 
{\em observables} again: as an example, 
it has been shown explicitly in the
Schwinger model that the axial anomaly is reproduced correctly, if we
map the entire continuum theory in a consistently perfect way on the 
lattice \cite{Schwing}. 
This includes the blocking of the fermion and gauge field, of the
fermion-gauge interaction term to the first order,
and of the axial current.
A perfect lattice current is identified by a procedure analogous
to the treatment of the fields.
The blocking scheme here integrates the continuum flux through
the face between adjacent lattice cells, and a coupling to an external
source incorporates the current in the RGT.
\footnote{The perfect currents, which have been worked out 
in this context, gave also rise to a study of perfectly discretized
hydrodynamics \cite{Katz}.}
The axial charge is blocked from the continuum too, and all these 
perfect quantities match to reproduce the correct axial anomaly
on the lattice.

By construction, the chiral symmetry comes out correctly to all orders
in perturbation theory, where the blocking from the continuum
can be carried out. So we can put the system on the 
lattice without doing any harm to it, i.e. the lattice regularization
is -- with this respect -- as good as any regularization in the 
continuum. (This is in contrast to a wide-spread feeling that it has
a particular weakness due to the fermionic doubling problem.)

The construction of {\em local} chiral fermions we sketched above
-- the fixed point action for $\alpha >0$ --
is a nice way around the Nielsen Ninomiya theorem, which refers
to a chiral symmetry of the lattice action itself. We don't need it to
be manifest in the action, 
\footnote{The violation in the form of the action is due to the
non chirally symmetric transformation term for $\alpha >0$.}
but we can still reproduce it in the 
observables, which really matter (like the continuous rotation and
translation symmetry, which we recovered in the spectrum).
However, since this construction by ``blocking from the continuum''
is perturbative, it does not directly imply any
claim about a non-perturbative formulation of the lattice chiral
fermion (which is not known in the continuum either).

While these notes are being written down, however,
it has been shown that also the Atyiah Singer index theorem
holds for a FPA \cite{Index}, confirming again that the perfect 
fermion is perfect.

\subsection{Truncation}

The perfect actions mentioned above all include
couplings over infinite distances. Even if the long range couplings
are exponentially suppressed (locality), they are still needed
to reproduce the continuum symmetries exactly.
Since we can't work with them in simulations, the nice properties
described above may seem somewhat academic,
rather far from application (is it comparable
to the string theory talks at this symposium ?).

To proceed to practical applications, we need to truncate the
couplings, and this does necessarily some harm to the perfect
properties. In the 2d applications, it was permissible to truncate 
only couplings, which are suppressed by various orders of magnitude
\cite{HN,Lang}.
However, in $d=4$ -- and with non-Abelian gauge fields,
where many more lattice paths have to be distinguished --
we can not be so generous. The number of additional terms, that
the practitioner can work with, is strongly limited.

Hence it is very important to choose the RGT such that the perfect
action is {\em as local as possible}, i.e. the couplings decay as
fast as possible. Then there is hope that perfection
survives the truncation in a good approximation.
In this sense, we have tuned the RGT parameter $\alpha$
in the RGT (\ref{fermiRGT}) in order to optimize locality
and we could achieve that in the special case
$p=(p_{1},0\dots 0)$ (mapping on $d=1$) 
the couplings are restricted to  nearest neighbors. 
This is  a successful optimization criterion for the
locality in the general 4d case, for fermions as 
well as scalars \cite{QuaGlu,WB}.
For the gauge field we have chosen the RGT parameter,
which is analogous to $\alpha$, to be momentum dependent.
Thus we obtained the standard plaquette action in $d=2$, 
and again the same RGT parameters also provide excellent 
locality in $d=4$ \cite{QuaGlu}.

\begin{figure}[hbt]
   \begin{tabular}{cc}
      \hspace{-0.8cm}
\def\fpsangle{0} \epsfxsize=80mm \fpsbox{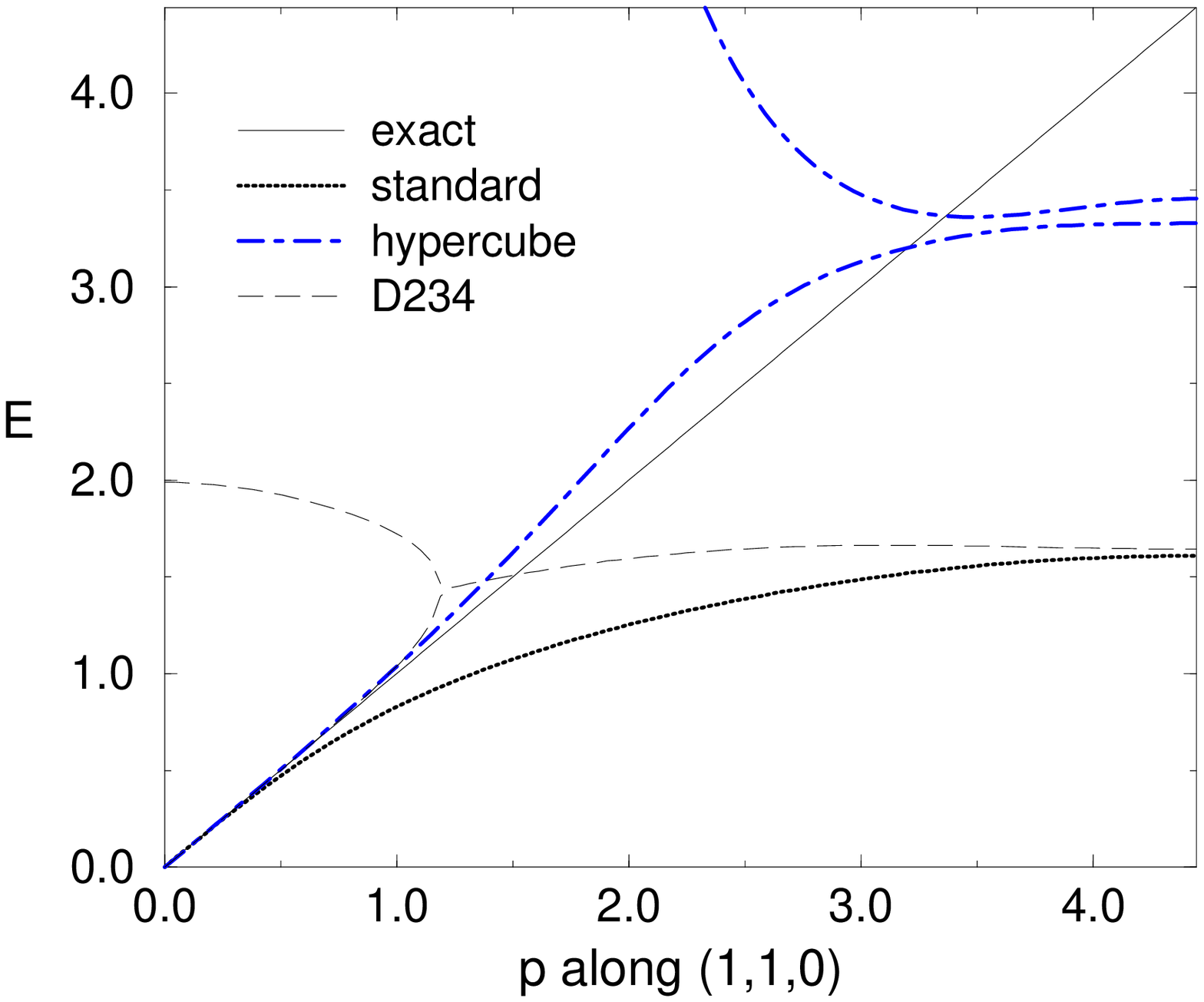} &
      \hspace{-1.0cm}
\def\fpsangle{270} \epsfxsize=67mm \fpsbox{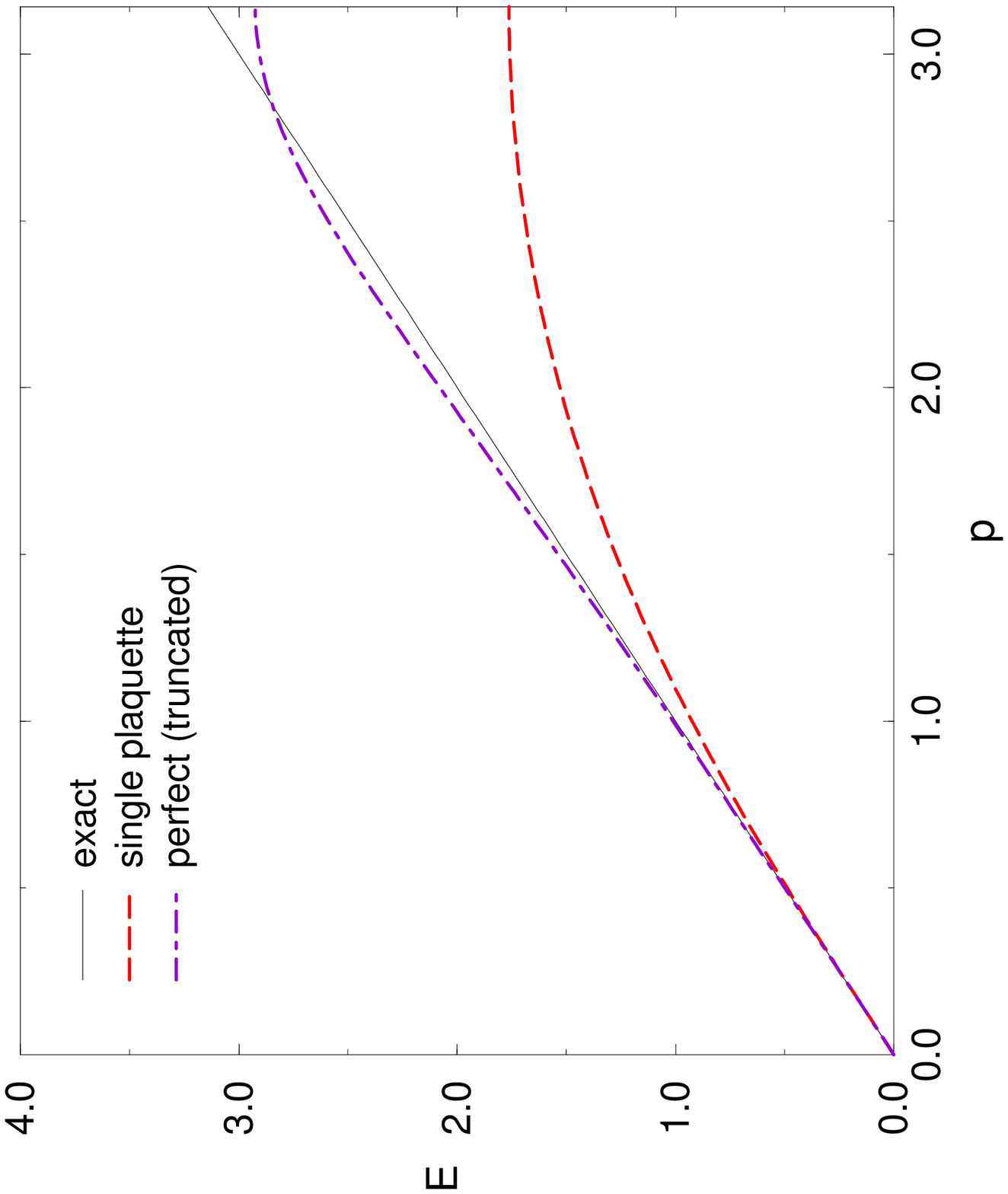}
   \end{tabular}
   \caption{\footnotesize
            {The dispersion relation for free massless fermions (left)
and gauge fields (right), compared to the Wilson action and --
for the fermion -- to a Symanzik improved action called D234.}}
   \label{spectra}
\end{figure}

\begin{figure}[hbt]
   \begin{tabular}{cc}
      \hspace{-0.5cm}
\def\fpsangle{0} \epsfxsize=75mm \fpsbox{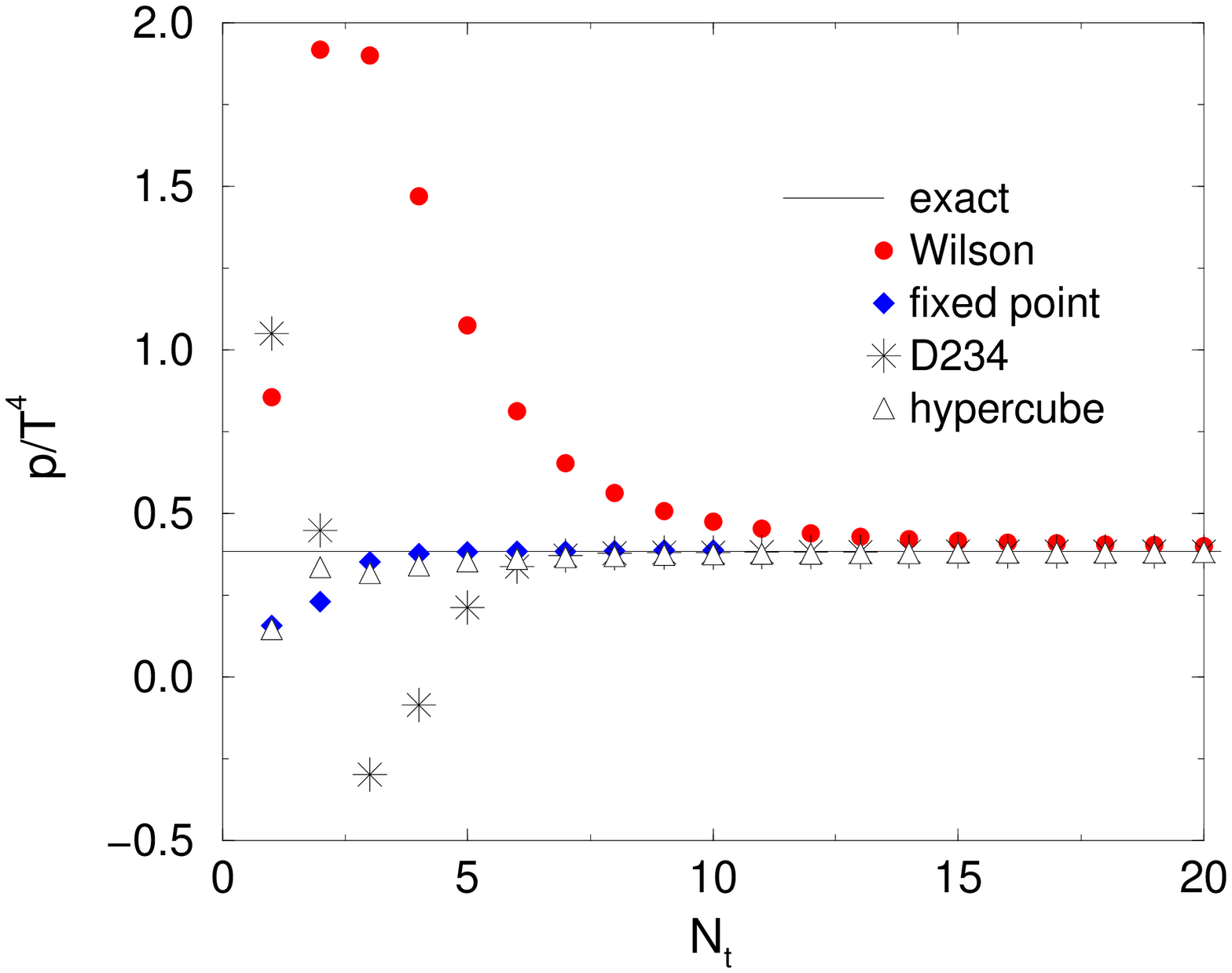} &
      \hspace{-1.2cm}
\def\fpsangle{270} \epsfxsize=58mm \fpsbox{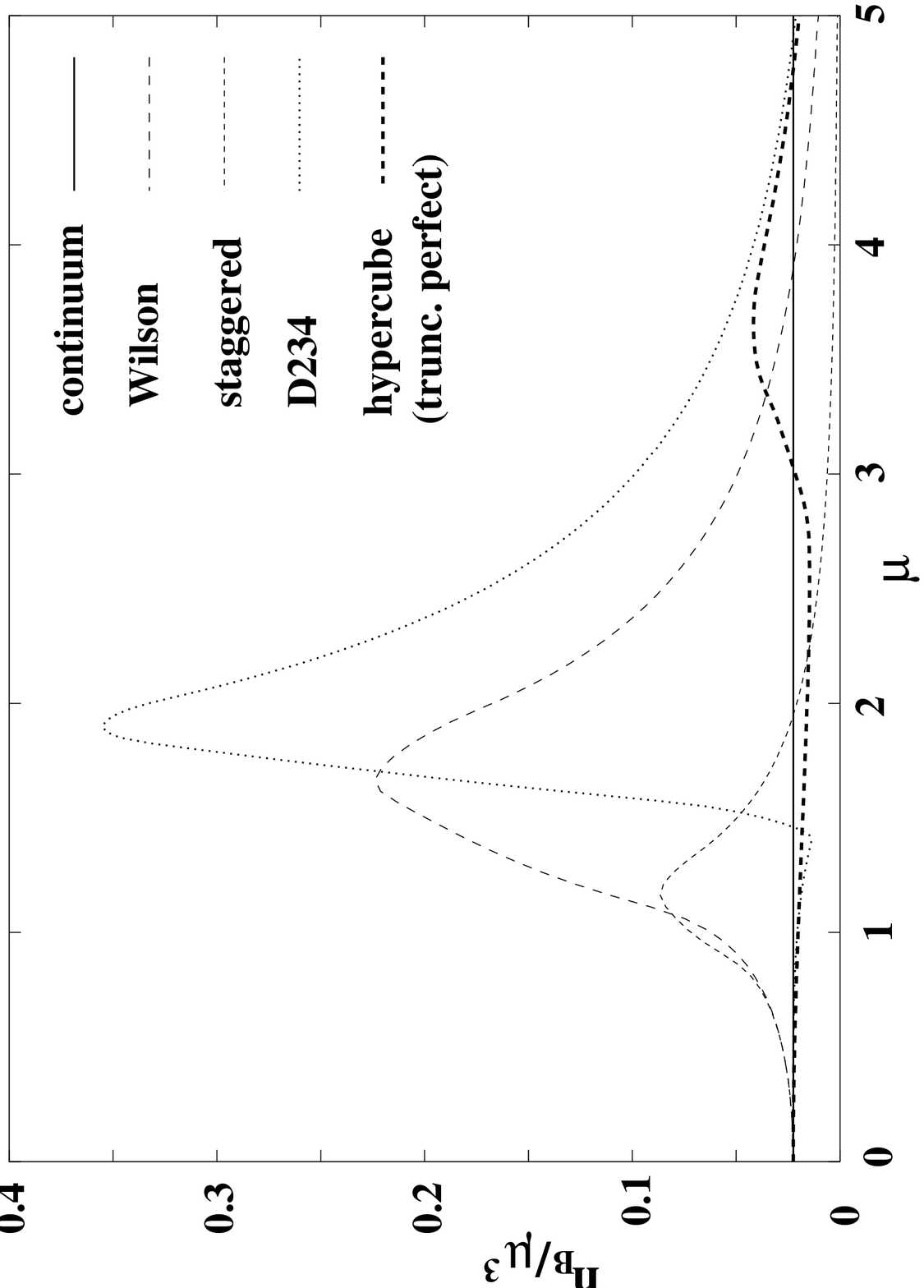}
   \end{tabular}
   \caption{\footnotesize
            {The thermodynamic scaling quantities 
$pressure/temperature^{4}$ with $N_t$ sites in Euclidean time
(at $\mu =0$), and 
$(baryon~number~density)/(chemical ~potential)^{3}$} at $T=0$,
for massless free fermions.}
   \label{thermo}
\end{figure}

The truncation itself was performed by means of periodic boundary 
conditions: we construct a perfect action in a small volume
of $3^{4}$ sites, and then use the same couplings 
-- which are restricted to a unit hypercube -- in larger
volumes too. This truncation keeps all normalizations exact.
To check the quality of the truncated free actions,
we considered the spectrum after truncation, which is not
perfect any more, but still drastically improved compared to the
standard action \cite{StL}, see Fig. \ref{spectra}.
The same holds for thermodynamic scaling quantities
at finite temperature \cite{StL,FStL} and at finite 
chemical potential \cite{chem}, as illustrated in Fig. 
\ref{thermo}.
A different truncation of the perfect free gauge field in terms 
of closed loops has been given in Ref. \cite{StL}.

\begin{figure}[hbt]
\hspace{3.5cm}
      \epsfxsize=8.00cm \epsfbox{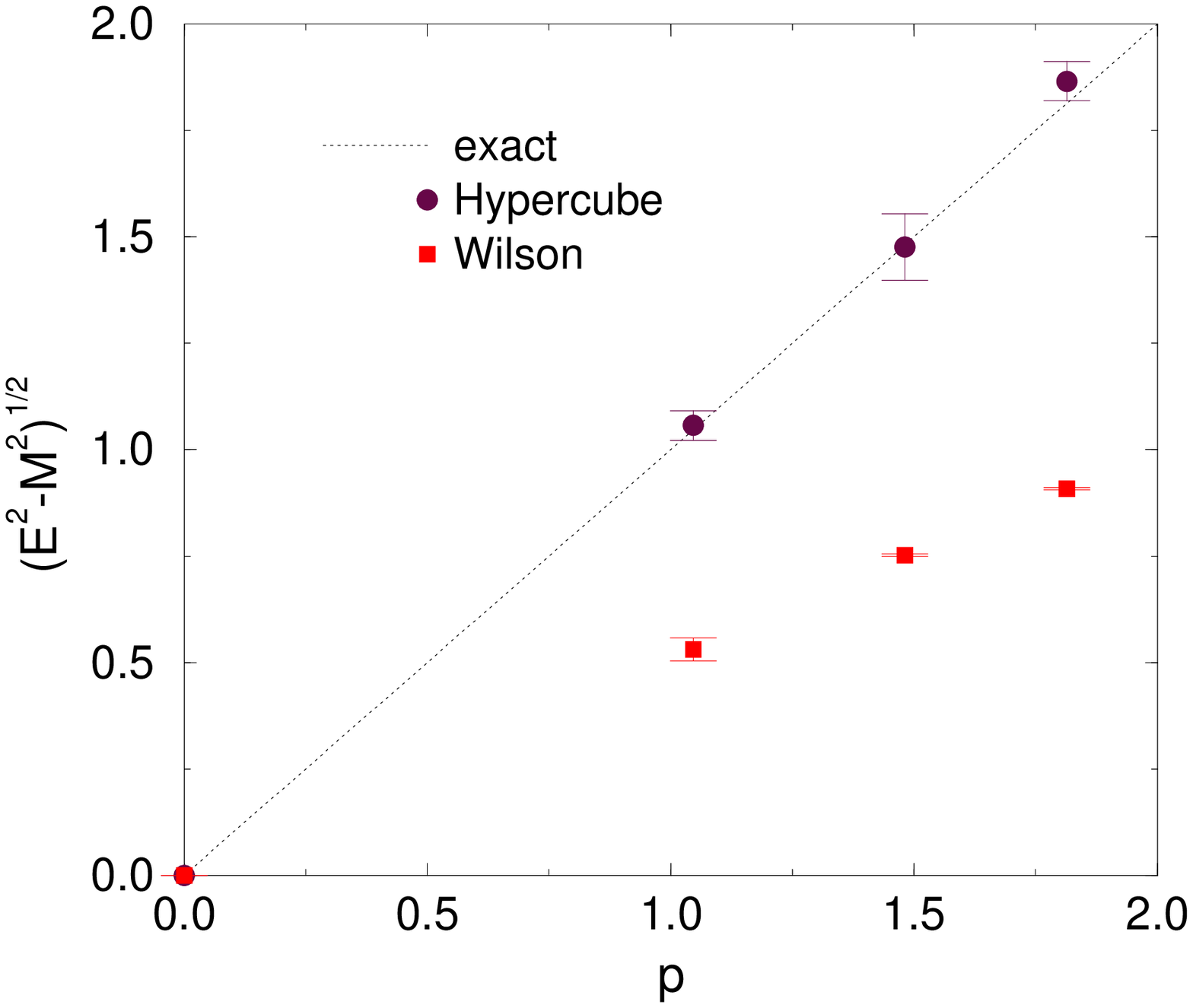}
\vspace{-5mm}
   \caption{\footnotesize
            {The meson dispersion relation using the truncated perfect
``hypercube fermion'', and the Wilson fermion.}}
   \label{mesodisp}
\end{figure}

For an immediate application, the truncated perfect
``hypercube fermion'' can be ``gauged by hand'':
we connect the coupled sites by the link variables
on the shortest lattice paths, and average over these paths.
This is of course a drastic short-cut and not the consistently 
perfect construction, but with some respect it already leads 
to a remarkable progress. In particular, the meson dispersion relation
reaches an impressive quality, as shown in Fig. \ref{mesodisp}.
To handle such complicated actions efficiently, work is in
progress for an adequate optimization of the algorithm
on a parallel machine.
In particular, we are working on a quick evaluation of the
fermion matrix \cite{Eick}, which involves many more non-zero off-diagonal
elements than it is the case for Wilson's action.

\subsection{Vertex function}

As we proceed to perturbatively interacting fields, we can incorporate
the interaction into the RGT and still block from the continuum.
If we realize this to the first order in the gauge coupling
$g$, we 
obtain a perfect quark-gluon
and a 3-gluon vertex function. The $\bar q q g$ vertex has been
evaluated in $d=2$ \cite{StL} and its couplings have been reproduced
numerically from a multigrid \cite{Lang}. In spite of good locality, we
obtain in the 4d case still inconveniently many couplings, 
which seem to contribute significantly.
By means of a rather tough truncation, their number was reduced
to 5 short ranged extra terms, in addition to those present in
the action of the ``hypercube fermion gauged by hand'' \cite{Org}.

\begin{figure}[hbt]
   \begin{tabular}{cc}
      \epsfxsize=7.5cm \epsfbox{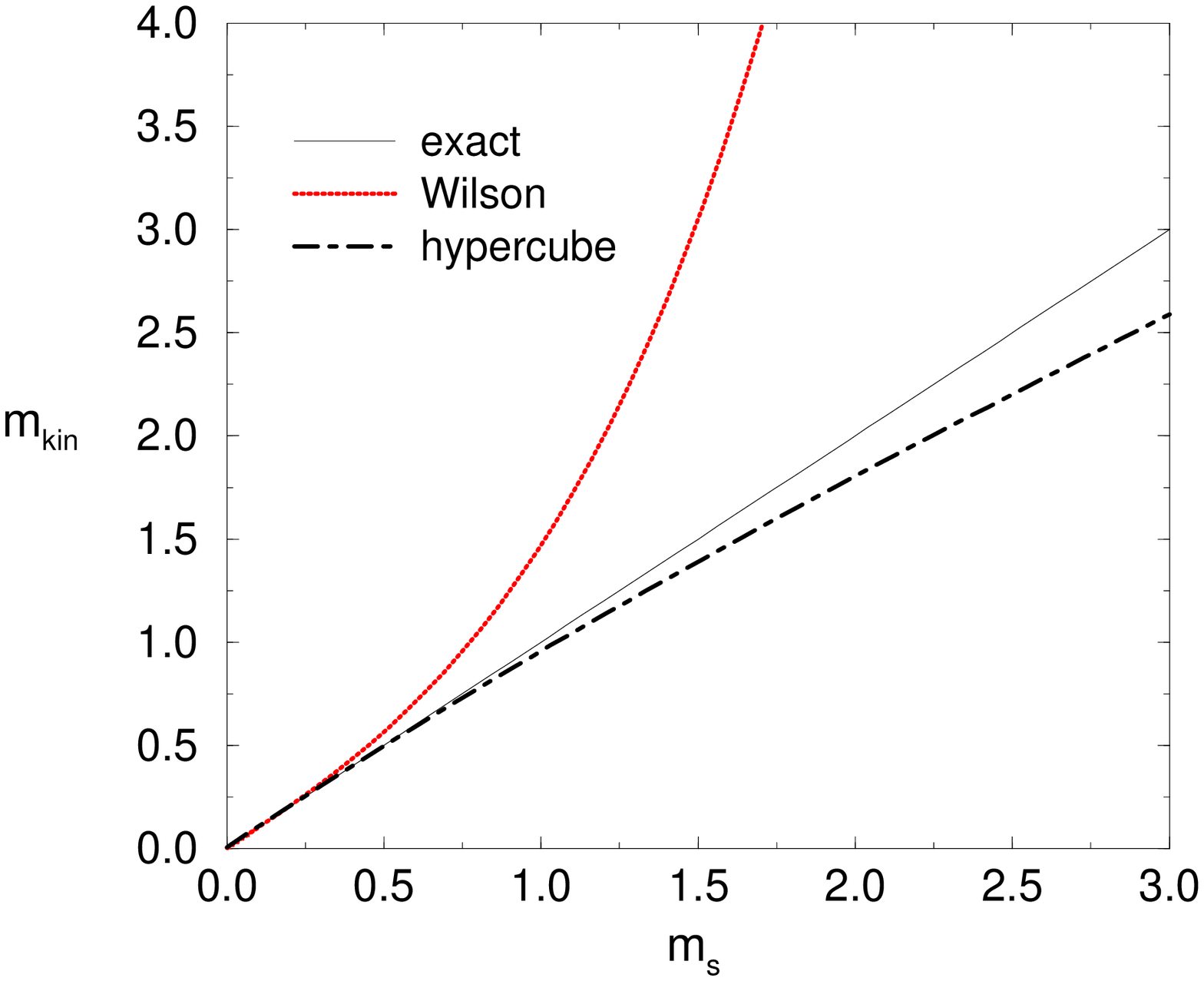}  &
      \epsfxsize=7.5cm \epsfbox{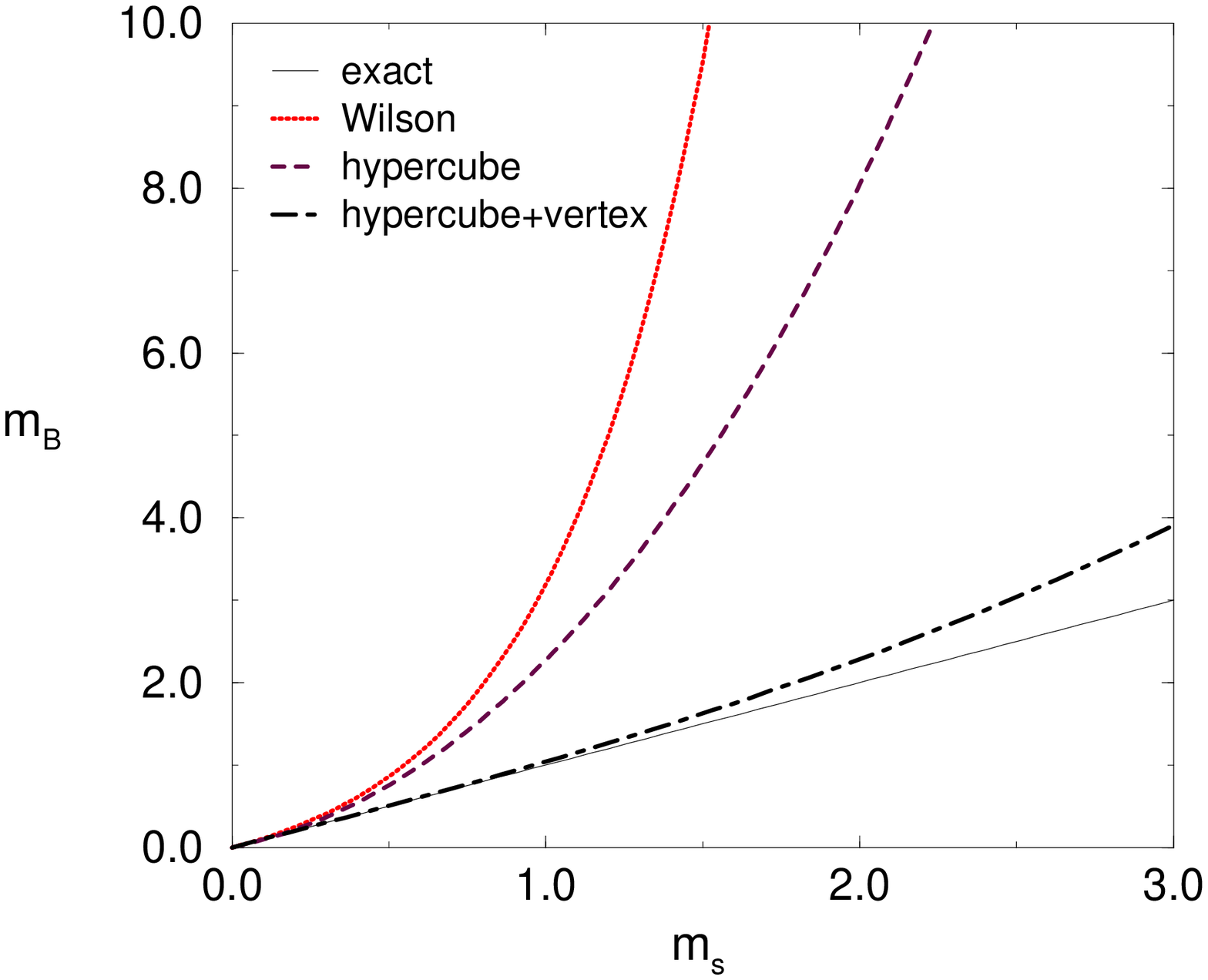}
   \end{tabular}
\vspace{-3mm}
   \caption{\footnotesize
            {The kinetic and the magnetic mass as functions of the
static mass for the Wilson fermion and the hypercube fermion.
$m_{B}$ improves strongly only after including elements of the
perfect vertex function.}}
   \label{nonrel}
\end{figure}

The best candidate for an immediate use of this action
is heavy quark physics.
This optimism is supported from a consideration of the 
non-relativistic expansion
\begin{equation}
E = m_{s} + \frac{1}{2m_{kin}} \vec p^{~2} + \frac{1}{2m_{B}}
\vec \Sigma \vec B + \dots \ , \quad
\Sigma_{k} = \epsilon_{ijk} \sigma_{ij}/2 .
\end{equation}
For the periodically truncated perfect fermion, the mass
parameter $m$ coincides with the static mass $m_{s}$.
The lattice parameters tend to deviate from the continuum
relation $m_{s}=m_{kin}=m_{B}$. Fig. \ref{nonrel} shows that
$m_{s}=m_{kin}$ is approximated well for the hypercube fermion,
and if we gauge it by hand then $m_B(m_{s})$ is somewhat 
improved. We can achieve a drastic improvement also for $m_{B}$ 
by including terms of the truncated perfect vertex function.

\begin{figure}[hbt]
\hspace{3.5cm}
      \epsfxsize=5cm \epsfbox{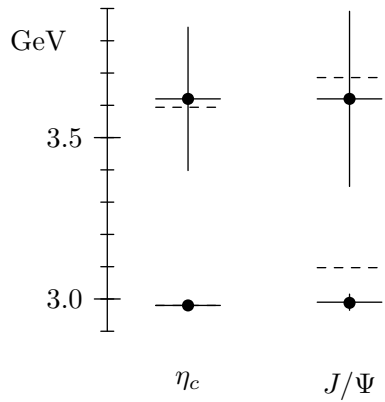}
\vspace{-2mm}
   \caption{\footnotesize
            {The charmonium spectrum. The experimental values are dashed,
and only the ground state of $\eta_{c}$ is fitted.}}
   \label{chaspe}
\end{figure}

As a first experiment, we performed quenched simulations with the 
truncated perfect vertex function
for the charmonium spectrum. The computational
overhead (compared to Wilson's action) is around a factor 20,
but we simulated on an coarse lattice of spacing $a=0.24 fm$
(about 3 times larger than usual), which could clearly out-do 
this factor. Here a lattice of $8^{3}\times 16$ sites appears
to be sufficient.
The physical units were determined from the string tension, and
the 1s $\eta_{c}$ ground state was matched to the experimental value.
The further states were predictions of the simulation.
The 2s states of $\eta_{c}$ and $J/\psi$ were predicted 
successfully, but the gap to the 1s $J/\psi$ state (hyperfine
splitting) was clearly too small \cite{Org}, see Fig. \ref{chaspe}.
A possible reason -- except for quenching -- is that
the vertex couplings are still supposed to be renormalized
(due to $O(g^{2})$).
As a rather ad hoc procedure, we have tested a ``tadpole
improvement'' (mean field estimate of the renormalization
\`{a} la Ref. \cite{LepMac}), which helps to some extent, but the
hyperfine splitting -- a quantity, which is extremely sensitive
to lattice artifacts -- is still too small.
However, it is known from the clover action
that the fermion part tends to be insufficiently
renormalized by this method, so we are now incorporating
the larger renormalization factor, translated 
from the result found by the ALPHA collaboration
for the clover term \cite{ALPHA}.

The form of the perfect $ggg$ vertex function is very similar to the 
$\bar q q g$ vertex \cite{KosDis}. 
If we include it, then we obtain an action
which is entirely perfect (before truncation) to $O(g)$;
any artifacts of $O(a^{n})$ and $O(ga^{n})$ are eliminated,
so that the leading artifacts occur in $O(g^{2}a)$.
It is not obvious how this compares to a Symanzik improved action,
which erases all artifacts in $O(ag^{n})$, but which is plagued
for instance by cut-off effects in $O(ga^{2})$.

The climax of the improvement is of course a QCD action, 
which is improved non-perturbatively
in both, $a$ and $g$. The classically perfect action, to be 
constructed from a multigrid minimization, is an action
of this type, but it could not be worked out in an
``ultimate'' form yet. In practice one would perform
inverse blocking steps with a small blocking factor
(see footnote 3),
but iteration rapidly leads to large lattices,
which can not be handled any more. Of course one could
put a well identified action, say after a blocking factor
2 RGT, back to the fine lattice and block again, hence avoiding
the use of large lattices. However, small lattices
are only instructive if the action is extremely local.
In any case, a good starting point for the iteration,
as well as a good parameterization ansatz for the blocked  action,
are highly desirable, and the perfect vertex functions could help
with this respect.

\section{Status and outlook}

Although we are still working on better charmonium results, 
it seems that the direct application of small field perfect 
actions is not as successful as we hoped. 
This is also confirmed by a study of
the anharmonic oscillator \cite{anos}. 
Apparently, at some point the minimization trick must come
into play. In simple cases one might think about extending
it to a semi-classically perfect action \`{a} la WKB, 
i.e. taking the quadratic fluctuations around the classical 
solution into account.
Also one full RGT step with a small blocking factor,
starting from a classically perfect action at moderate $\xi$,
could be very helpful.

One point that is still very important to work on,
is the optimization of locality: in the case of the vertex 
functions, it seems that the interaction term in the RGT 
can be further optimized with this respect.
In other cases, e.g. for so-called staggered fermions 
(see second and third Ref. in \cite{books}), it turned out
that one can do better even for the free particles by taking
a blocking scheme different from the block average described
in Sec. 2 \cite{Dilg}. The fine lattice (or continuum) variables can be
blocked in many ways, not only by a piece-wise constant
weight factor as in eq. (\ref{blocking}). 
Such an optimization led to significant progress for the
truncated perfect staggered fermion, which has been tested in
simulations of the Schwinger model 
(with naive gauging plus ``fat links'');
in particular the ``pion'' mass gets strongly suppressed.

In thermodynamics, where lattice artifacts are especially bad,
a remarkable improvement has been observed using approximately
perfect actions \cite{thermo}. 
As a further step, one could put those
actions on anisotropic lattices, which hardly poses additional
complications \cite{WB}. 
It has also been shown how
to include a chemical potential in a perfect action, which
is needed for simulations at finite baryon density \cite{chem}
(a field, which had only little success so far).

In further experiments for QCD, pionic systems were investigated
and some improvement in the scaling behavior was found \cite{TdG}, 
but that study also tends to involve more and more ad hoc elements,
which are not related to the perfect program. From that side,
there is no claim for an ``ultimate'' version of a 
quasi-perfect action either.
A problem, which our attempts have in common, is a strong
additive mass renormalization. This is a effect of truncation
(and simplified gauging),
which affects the reliability of the ``perfect couplings''.

According to our experience,
it is worthwhile proceeding in very small steps to higher degrees
of complications, in order to elaborate a really clean treatment.
All attempts at short-cuts to QCD have led to half-cooked 
proposals, which have to be reconsidered afterwards.
Thus the program is quite time-consuming, but if it really
works one day, a quasi-perfect action (given by a set of couplings)
can be used by the practitioner for any problem in QCD, without 
worrying about its lengthy derivation.
Following these piecemeal tactics, we are now working on a
FPA for 2d and 3d non-Abelian gauge theories, starting with
$SU(2)$. As a scaling test, the 3d glueball spectrum can be 
measured.

\begin{figure}[hbt]
\vspace{-0.7cm}
\hspace*{0.5cm}
\def\fpsangle{270}
\epsfxsize=100mm
\fpsbox{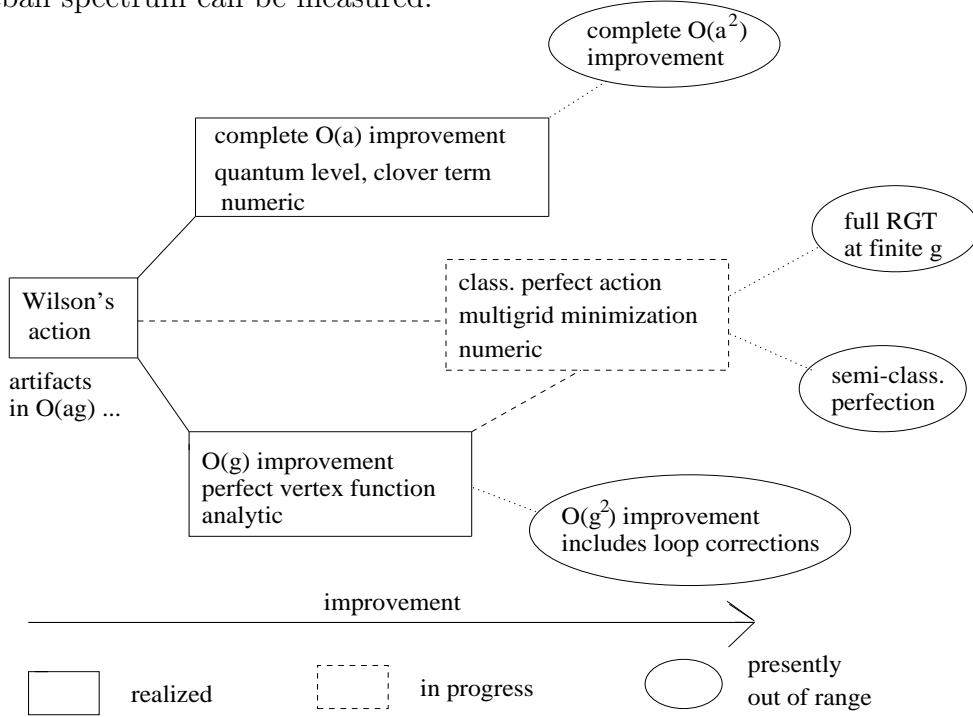}
\vspace{-0.2cm}
  \caption{\footnotesize
            {A schematic overview over the status of the
improvement for QCD lattice actions with Wilson-type fermions.}}
   \label{schema}
\end{figure}

On the conceptual side, it is straightforward to apply
the procedure of blocking from the continuum for instance
to supersymmetry. This results in a SUSY lattice action, which is 
completely different from the one presented by I. Montvay
at this symposium. Again it is more complicated, but it could
represent all symmetries exactly. Even applications to quantum
gravity are conceivable: also here it would be desirable to integrate
out the short-range details between the lattice sites in a way,
which is based on the renormalization group. Again, such a method
would differ from those discussed by J. Ambj/$\!\!\! {\rm o}$rn and
B. Pertersson in their talks.

At last, as another project in progress,
it would be interesting to combine
the perfect treatment of fermions with the domain wall
concepts of lattice chiral fermions.
So far, that construction has always been based on Wilson
fermions \cite{dom}, although this can be generalized.
Inserting a truncated perfect fermion instead, one cumulates
all sort of virtues: small lattice artifacts, an arbitrary
number of flavors (unlike staggered fermions), and the absence
of additive mass renormalization (unlike Wilson fermions).
Note that the domain wall formalism for chiral fermions
is related to the subject, which was most celebrated at this symposium,
the theory of D-branes (I thank J. Schwarz for this remark). \\

\vspace*{-1mm}
{\em Acknowledgment} 
It is a pleasure to thank R. Brower,
S. Chandrasekharan, H. Dilger, E. Focht, K. Orgions, T. Struckmann,
U.-J. Wiese and the co-authors of Ref. \cite{Eick}
for their collaboration.

\end{document}